# Carbon Monitor Europe, near-real-time daily $CO_2$ emissions for 27 EU countries and the United Kingdom

Piyu Ke[1], Zhu Deng[1], Biqing Zhu[1], Bo Zheng[2], Yilong Wang[3], Olivier Boucher[4], Simon Ben Arous[5], Chuanlong Zhou[6], Xinyu Dou[1], Taochun Sun[1], Zhao Li[1], Feifan Yan[7], Duo Cui[1], Yifan Hu[8], Da Huo[1], Jean Pierre[9], Richard Engelen[10], Steven J. Davis[11], Philippe Ciais[6,*], Zhu Liu[1,*]

[1] *Department of Earth System Science, Tsinghua University, Beijing, China*
[2] *Tsinghua Shenzhen International Graduate School, Tsinghua University, China*
[3] *Key Laboratory of Land Surface Pattern and Simulation, Institute of Geographical Sciences and Natural Resources Research, Chinese Academy of Sciences, Beijing, 100101, China*
[4] *Institute Pierre-Simon Laplace, Sorbonne Université/CNRS, Paris, France*
[5] *Kayrros, 33 Rue La Fayette, 75009 Paris, France*
[6] *Laboratoire des Sciences du Climate et de l'Environnement LSCE, Orme de Merisiers 91191 Gif-sur-Yvette, France*
[7] *Key Laboratory of Marine Environment and Ecology, and Frontiers Science Center for Deep Ocean Multispheres and Earth System, Ministry of Education, Ocean University of China, Qingdao 266100, China*
[8] *Key Laboratory of Sustainable Forest Ecosystem Management, Northeast Forestry University, Harbin 150040, China*
[9] *CITEPA, 42 Rue de Paradis, 75010 Paris*
[10] *European Centre for Medium-Range Weather Forecasts, Reading, RG2 9AX, UK*
[11] *Department of Earth System Science, University of California, Irvine, 3232 Croul Hall, Irvine, CA, 92697-3100, USA*

[*]Corresponding authors: Zhu Liu (zhuliu@tsinghua.edu.cn), Philippe Ciais (philippe.ciais@cea.fr)

## Abstract

With the urgent need to implement the EU countries pledges and to monitor the effectiveness of Green Deal plan, Monitoring Reporting and Verification tools are needed to track how emissions are changing for all the sectors. Current official inventories only provide annual estimates of national $CO_2$ emissions with a lag of 1+ year which do not capture the variations of emissions due to recent shocks including COVID lockdowns and economic rebounds, war in Ukraine. Here we present a near-real-time country-level dataset of daily fossil fuel and cement emissions from January 2019 through December 2021 for 27 EU countries and UK, which called Carbon Monitor Europe. The data are calculated separately for six sectors: power, industry, ground transportation, domestic aviation, international aviation and residential. Daily $CO_2$ emissions are estimated from a large set of activity data compiled from different sources. The goal of this dataset is to improve the timeliness and temporal resolution of emissions for European countries, to inform the public and decision makers about current emissions changes in Europe.

## Background & Summary

The European Union and the United Kingdom is the world's third energy consumer and $CO_2$ emitter, accounting for 12% of global emissions [1-3]. The EU27 announced the Green Deal in December 2019, and set up a road map for cutting greenhouse gas emissions by at least 55% by 2030 and reaching carbon neutrality by 2050[4]. With increasing focus and effort on reducing $CO_2$ emissions, there is a growing need for more reliable data as well as for data with a lower latency. Most $CO_2$ emission inventories including national inventories reported to the United Nations Framework Convention on Climate Change (UNFCCC) lag reality by one years or more[2,5-10].

Lower latency estimates of $CO_2$ emissions are generally delayed by several months as data must be gathered from numerous sources and then verified. For example, Eurostat has been producing early estimates of annual and country-level $CO_2$ emissions in the EU since at least 2012, with a delay for about 5 months[11]. The Global Carbon Project publishes projections of the current year's global fossil $CO_2$ emissions since 2012 by the end of the current year[12]. As the COVID-19 pandemic profoundly disrupted human activities in the year 2020, existing inventories were not able to monitor changes in activity and assess the COVID-19 impacts on $CO_2$ emissions during that period and the following recovery. Le Quéré et al.[13] and Forster et al.[14] developed methods for estimating daily, global $CO_2$ emissions, based on confinement index and Google mobility indexes respectively[13,14]. These two approaches captured first order pandemic related reductions but were not tested for subsequent variations when lockdowns were finished, and did not continue to provide near-real-time, daily and country-level estimates emissions. Following the pandemics, few EU countries and the UK started to publish quarterly or monthly estimates of emissions from preliminary energy data[15-19]. Further, the private company Kayrros released a daily estimate of emissions from regulated sectors (European Trading Scheme) in November 2021 using site-level satellite monitoring of industrial activity, called Carbon Watch[20]. The Carbon Monitor international research initiative developed a new near-real-time daily dataset of $CO_2$ emissions with global coverage and country-level estimates for 12 countries which lags reality by only one month[1,3,21-25].

Here, we present a near-real-time, sector-specific, country-level estimates of daily fossil fuel and cement $CO_2$ emissions for 27 European Union countries and the United

Kingdom based on an extension of Carbon Monitor called Carbon Monitor Europe (CM-EU). We present the methodology and the results for changes in emissions between January 1, 2019 and December 31, 2021. Details of our data sources and approach are in the Methods section. An evaluation of CM-EU against national quarterly emissions estimates and Carbon Watch is provided in the technical validation section. The data Carbon Monitor Europe are publicly available at https://eu.carbonmonitor.org/.

## Methods

Carbon Monitor Europe (CM-EU) is a new regional improved version of the global Carbon Monitor system, providing near-real-time daily estimates of $CO_2$ emissions for six sectors over the globe with separate estimates for 12 countries or groups of countries [21]. CM-EU presents country-level estimates of daily $CO_2$ emissions from January 2019 through December 2021 for 27 countries of European Union (EU27) and the United Kingdom (UK). We used annual $CO_2$ emissions of EU27 & UK from the Emissions Database for Global Atmospheric Research (EDGAR) [8] as the baseline data for emissions in the year 2019, then disaggregated tis estimate into daily scale and make a projection to 2020 and 2021 based on time variations of activity data. Below we describe the calculation process in detail.

### Annual country-level and sectoral $CO_2$ emissions in the baseline year 2019

Annual fossil fuel combustion and cement production $CO_2$ emissions by sector in 2019 for all European Union countries and United Kingdom are directly obtained from the *Fossil $CO_2$ emissions of all world countries - 2020 Report* [8] released by the Emissions Database for Global Atmospheric Research (EDGAR). Emissions of EDGAR are derived using the IPCC Tier 1 approach according to the 2006 IPCC Guidelines [26], with 35 sectors based on IPCC categories. We aggregated 11 energy-related and cement production sectors of EDGAR into six main sectors, including power, industry, ground transportation, residential (public, commercial buildings and households), domestic and international aviation. By convention, emissions from international flights are assigned to the country of departure. Table 1 shows the correspondence table between aggregated sectors of CM-EU and EDGAR sectors based on IPCC (first column).

### Data acquisition and processing of daily $CO_2$ emissions in 2019, 2020 and 2021

Carbon Monitor follows the IPCC guidelines for emissions reporting [26] in computing $CO_2$ emissions from a country / region by multiplying activity data (AD) by corresponding emissions factors (EF):

$$Emis = \sum\sum\sum AD_{i,j,k} \times EF_{i,j,k}$$

where i, j, k denote regions, sectors, and fuel types respectively. We assume that emission factors and structure of each sector remain unchanged for each country in 2020 and 2021 compared with 2019. Thus, the rate of change of the emission is calculated based solely on the change of the activity data in 2020 and 2021 compared to the same period of 2019. The emissions were calculated following this equation separately for the power sector, the industrial sector, the ground transportation sector, the aviation sector (including the domestic and international aviation sector) and the residential sector.

Power sector

For the power sector, daily emissions were calculated from hourly electricity generation data  by production types at resolution of 1 h to 15 min. Data of 22 EU countries (Austria, Belgium, Bulgaria, Croatia, Cyprus, Czech Republic, Denmark, Estonia, Finland, France, Germany, Greece, Hungary, Ireland, Italy, Latvia, Netherlands, Poland, Portugal, Slovakia, Slovenia and Spain) and United Kingdom are available from the ENTSO-E Transparency platform (https://transparency.entsoe.eu/dashboard/show) and Balancing Mechanism Reporting Service (BMRS) (https://www.bmreports.com/) respectively (Table 2). We removed outliers and filled the N/A values by using the "interpolate()" function in Python Pandas packages, then aggregated the thermal production data into daily level. The electricity generation data used in this study from ENTSO-E and BMRS are include several fuel types with coal, gas, oil and peat. The average yearly emission factor of the whole power sector is assumed to remain constant at its 2019 value, so that daily emissions are estimated by:

$$Emis_{power,daily,2019} = Emis_{power,yearly,2019} \times \frac{AD_{power,daily,2019}}{AD_{power,yearly,2019}}$$

$$Emis_{power,daily,2020\ or\ 2021} = Emis_{power,daily,2019} \times \frac{AD_{power,daily,2020\ or\ 2021}}{AD_{power,daily,2019}}$$

For Lithuania, Luxembourg, Malta, Romania and Sweden (other EU) not covered by ENSTO-E, we assume a linear relationship between their daily emissions and the total daily emissions of the 22 EU countries and United Kingdom (EU22+UK) and their daily emissions are estimated as:

$$Emis_{power,daily,2019,EU5} = Emis_{power,yearly,2019,other\ EU} \times \frac{Emis_{power,daily,2019,EU22+UK}}{Emis_{power,yearly,2019,EU22+UK}}$$

$$Emis_{power,daily,2020\ or\ 2021,other\ EU} = Emis_{power,daily,2019,EU5} \times \frac{Emis_{power,daily,2020\ or\ 2021,EU22+UK}}{Emis_{power,daily,2019,EU22+UK}}$$

Industry sector

Daily emissions from industry are estimated using the monthly industrial production index (IPI) from several datasets and daily power generation data (Table 2). As daily production data are not available for industrial and cement production, the monthly $CO_2$ emissions are estimated by using monthly statistics of industrial production and daily data of electricity generation to disaggregate the monthly $CO_2$ emissions into a daily scale. This approach is based on two assumptions: 1. A linear relationship is assumed between industrial production index and emissions from industrial and cement production. 2. A linear relationship is assumed between daily industry activity and daily electricity production, from ENSTO-E and our approach for the five Other-EU countries. Therefore, the monthly and daily industry emissions are estimated following:

$$Emis_{industry,monthly,2019} = Emis_{industry,yearly,2019} \times \frac{AD_{industry,monthly,2019}}{AD_{industry,yearly,2019}}$$

$$Emis_{industry,monthly,2020\ or\ 2021} = Emis_{industry,monthly,2019} \times \frac{AD_{industry,monthly,2020\ or\ 2021}}{AD_{industry,monthly,2019}}$$

$$Emis_{industry,daily,2019,2020\ or\ 2021} = Emis_{industry,monthly,2019,2020\ or\ 2021} \times \frac{AD_{power,daily,2019,2020\ or\ 2021}}{AD_{power,monthly,2019,2020\ or\ 2021}}$$

Ground transportation sector

Carbon Monitor uses TomTom congestion global level data (Table 2) from the TomTom website (https://www.tomtom.com/en_gb/traffic-index/) to capture the daily variations in the ground transportation activity. The TomTom traffic congestion level (called G hereafter) represents the extra time spent on a trip in congested conditions, as a percentage, compared to uncongested conditions. TomTom congestion level data cover 203 cities across 24 EU countries (Austria, Belgium, Bulgaria, Czech Republic, Denmark, Estonia, Finland, France, Germany, Greece, Hungary, Ireland, Italy, Latvia, Lithuania, Luxembourg, Netherlands, Poland, Portugal, Romania, Slovakia, Slovenia, Spain, Sweden) and the United Kingdom at a temporal resolution of one hour (Table 3). Note that a zero-congestion level means the traffic is fluid or 'normal' but does not mean there are no vehicles and zero emissions. The lower threshold of emissions when the congestion level is zero was estimated using real-time data from an average of 70 main roads in the city of Paris. The daily mean car counts (called Q hereafter) were calculated by a sigmoid function based on regression:

$$Q = a + \frac{bG^c}{d^c + G^c}$$

Where a, b, c, d are regression parameters. Then we allocated the country-level emissions from EDGAR into a daily scale using:

$$Emis_{ground\ transport,d} = Emis_{ground\ transport,yearly} \times \frac{Q_d}{\sum_{d=1}^{n} Q_d}$$

where $Q_d$ is the mean vehicle number per hour in day d, $Emis_{ground\ transport,d}$ is the ground transport emission in day d, $Emis_{ground\ transport,yearly}$ is the annual road transportation emissions from EDGAR, n is the number of days in a year. For countries not covered by TomTom (Croatia, Cyprus, Malta), we assume that their emission changes follow the patterns of total daily emissions from the average other 24 EU countries and the United Kingdom.

Aviation sector

Emissions of the aviation sector separated into domestic and international aviation, are estimated by individual commercial flights data (Table 2) from the Flightradar24 database (https://www.flightradar24.com). We compute $CO_2$ emissions by assuming a constant emissions factor of aviation in $CO_2$ emissions per km flown, called $EF_{aviation}$ across the whole fleet of aircraft (regional, narrowbody passenger, widebody passenger and freight operations) as the share of flight types has not significantly changed since 2019. The aviation sector separates domestic flights and international flights departing from countries considered in this study. The flights within EU but different countries are considered as international. The daily emissions of the aviation sector are estimated as:

$$Emis_{aviation,daily} = DKF \times EF_{aviation}$$

where DKF is the distance flown that is computed using great circle distance between the take-off, cruising, descent and landing points for each flight and are cumulated over all flights. For countries with overseas territories (mainly France, UK, Denmark) aviation emissions of flights to / from overseas territories are counted in this study as international emissions, whereas they would be reported as domestic emissions by national estimates.

Residential sector

Carbon Monitor uses the fluctuation of air temperature (Table 2) to capture the daily variations in the energy consumption of residential and commercial buildings. In this approach, only temperature is assumed to cause variation of emissions. Ciais et al.[27] looked at changes of natural gas use in the residential sector for few European countries from ENSTO-G pipeline data during the first half of the year 2020 and found that temperature variations dominated the observed changes, excepted for France and Italy where a reduced consumption signal independent of temperature was found during the weeks of strict confinements. The calculation of residential emissions was performed in three steps: (1) Calculation of population-weighted heating degree days for each country of EU27 & UK and for each day based on the ERA5 reanalysis of 2-m air temperature [28,29], (2) Using annual residential emissions in 2019 from EDGAR as the baseline. For each country, the residential emissions were split into two parts, i.e., cooking emissions and heating emissions, according to the EDGAR guidelines. The emissions from cooking were assumed to remain stable, while the emissions from heating were assumed to depend on and vary by the heating demand. (3) Based on the change of population-weighted heating degree days in each country, we scaled the EDGAR 2019 residential emissions to 2020 and 2021. Since the index of heating degree days are daily values, we get daily emission updates for the residential sources. The EDGAR residential emissions are downscaled to daily values based on daily variations in population-weighted heating degree days as follows:

$$Emis_{residential,monthly} = Emis_{residential,yearly,2019} \times \frac{\sum_m HDD_{daily}}{\sum_{m,2019} HDD_{daily,2019}}$$

$$Emis_{residential,daily} = Emis_{residential,monthly} \times Ratio_{heating,monthly} \times \frac{HDD_{daily}}{\sum_m HDD_{daily}} + Emis_{residential,monthly} \times (1 - Ratio_{heating,monthly}) \times \frac{1}{N_m}$$

$$HDD_{daily} = \frac{\sum(Pop_{grid} \times (T_{grid,daily} - 18))}{\sum Pop_{grid}}$$

Where $m$ is the month, $HDD_{daily}$ the population-weighted heating degree day, $Ratio_{heating,monthly}$ the percentage of residential emissions from heating demand monthly, $N_m$ the number of days in month m, $Pop_{grid}$ the gridded population data derived from Gridded Population of the World, Version 4 [28], $T_{grid,daily}$ is the daily average air temperature at 2 meters derived from ERA5 [29], and 18 is a HDD reference temperature of 18 °C following ref. [30].

## Data Records

The CM-EU dataset is an CSV file containing country-level emission of 27 European Union countries and the United Kingdom from 01/01/2019 to 31/12/2021 for six sectors: power, industry, ground transportation, domestic and international aviation. At the time of writing this article, this dataset has been updated to December 31, 2021, and the full dataset can be downloaded at Figshare[31]. Latest updates and related information are available for view and download on our website https://eu.carbonmonitor.org/.

Figure 1 presents daily $CO_2$ emissions in 2019, 2020 and 2021 for EU27 & UK, Germany, the Kingdom, Italy, Poland, France, Spain, Netherlands and the rest of European countries. As a whole, EU27 & UK emissions decreased by 10.36% (337.56 Mt $CO_2$) in 2020 as compared to 2019, then rebounded by 8.82% (256.96 Mt $CO_2$) in 2021 compared to 2020. EU27 & UK emissions in 2021 did not recover to the level of 2019, with a decrease by 2.71% (88.48 Mt $CO_2$) compared with 2019. The sudden and sharp drop of emissions in March and April of 2020 corresponds to the start of strict COVID-19 lockdowns. The COVID-related decrease in 2020 emissions was most pronounced in April and May (−28.13% during those months compared with the same period in 2019). Emissions began to recover in late May and June 2020, as lockdown restrictions in many European countries eased. The bars beneath each panel show the sectoral breakdown of annual emissions in 2021.

Among countries, the emission declines in 2020 and rebounds in 2021 varied considerably as shown in Fig. 1. The emissions in Germany decreased most in 2020 (-65.5 Mt $CO_2$) and rebounded most in 2021 (+53.5 Mt $CO_2$) (Figure 3). The country with the largest relative decrease in 2020 and increase in 2021 is Estonia, with -30% and +34% respectively (Figure 4). Lithuania is the only country where emissions did not fall in 2020 (+2.4%) and continued to grow in 2021 (+8.5%) (Figure 4).

Regrading sectors, the emission declines in 2020 and rebounds in 2021 of EU27 & UK show a change in the sectoral structure of the emissions (Figure 2 and Figure 3). Most sectors had significant reductions in emissions by 2020, especially the international aviation sector (-111 Mt $CO_2$). Emissions from each sector gradually recovered to pre-pandemic levels, as lockdown restrictions eased. However, the emissions of the domestic and international aviation in 2021 did not recover to the level of 2019 and remained lower by 31.7% (-3.81 Mt $CO_2$) and 48.3% (-91.16 Mt $CO_2$) compared with their 2019 level, respectively.

## Technical Validation

### Comparison with national emissions estimates, yearly and sub-yearly in selected countries, and Kayrros Carbon Watch data

To evaluate the $CO_2$ emissions data from CM-EU, we compared our results with three different existing emissions estimates of various timescale. The first evaluation data are annual country-level $CO_2$ emissions of EU27 & UK in 2019 and 2020 from EDGAR[8], BP[7], IEA[32], GCP[2] and Eurostat[33]. The second evaluation data are preliminary emissions publicly disclosed on a monthly or quarterly scale with a latency of few months to one year by national statistics offices / inventory agencies of Netherlands, Sweden, UK, France and Germany, which are from Centraal Bureau van Statistiek[18], Statistics Sweden[19], gov.uk[17], Citepa[16] and Umwelt Bundesamt[15] respectively. The third evaluation data are daily country-level $CO_2$ emissions of EU countries and UK from 2019 to 2021 from the Carbon Watch data of Kayrros [20] focusing on regulated sectors (ETS). As the Carbon Watch data methodology is not fully published, we provide a description as follows. There are four sectors in the data of Carbon Watch, including domestic flight, power, industry and transport.

*Domestic Flights*. Kayrros Carbon Watch computes the emissions from all the flights inside the European Economic Area, excluding the outermost regions. For each flight, the emissions are computed depending on the aircraft type and an estimation of the passenger load factor, as well as the distance flown[34]. The emissions are then aggregated by operators (the company responsible for the flight). Each company is then associated with a country based on its head office's location. If the head office of the operator is not located, it is attributed to the country where it operates the most, as per the ETS accounting methodology. Thus, these emissions are not 'territorial' like those of CM-EU based on the country from which planes take-off. Carbon Watch also includes emissions from flights to overseas territories as 'domestic' whereas they are considered as 'international' by Carbon Monitor.

*Power*. Kayrros Carbon Watch computes the emissions of the power sector using country-wise figures for electricity production by fuel type from the same data sources as CM-EU. Then an emission factor based on the fuel type and average plant's efficiency is used, based on the ETS register of CO2 emissions by each plant whereas CM-EU uses and aggregated emission factor for all the plants using the same fuel type.

*Industry*. Kayrros Carbon Watch computes the emissions of this sector by detecting activity signals from satellites and multiplying the activity by an emission factor. The activity signal of each site during cloud-free periods is detected using Sentinel 2 satellite images, interpolated in time for missing data, and aggregated for all sites of the same sub sector (e.g. all the cement plants). Subsectors currently not measured via satellites are modeled. In total, Kayrros Carbon Watch coverage currently includes the full scope of regulated industrial installations.

*Ground transport*. Kayrros Carbon Watch computes Ground Transport emissions through geolocation data from cellular phones, with a coverage of few percent of the population in EU countries. It computes the total distance traveled by all high-quality users (people with more than 20 daily pings). Then this proxy is converted to emissions based on IEA total ground transportation. CO2 emissions data[35] using an average emission factor by country based on the local split of gasoline and diesel. The results of all validation datasets are shown in Table 4.

Figure 5 shows the comparison of annual $CO_2$ emissions in 2019 and 2020 and changes between the two years for EU27 & UK, Germany, UK, Italy, Poland, France, Spain, Netherlands and the rest of EU between CM-EU, EDGAR, BP, IEA, GCP and Eurostat. In terms of annual total amount, the results of the six databases are relatively similar, and the main difference comes from the scope difference of each dataset. As for changes between 2019 and 2020, our estimates are lying in the middle range, close to most other datasets. Figure 6 shows the comparison of annual or quarterly $CO_2$ emissions in 2019 and changes between 2020 and 2019 or 2021 and 2020 for Netherlands, UK, Sweden and Germany between this study and estimates from Citepa (France), Centraal Bureau van Statistiek (Netherlands), gov.uk (UK), Statistics

Sweden (Sweden) and Umwelt Bundesamt (Germany). It illustrates that our estimates are close to the official estimates in terms of annual or quarterly totals for 2019 and for changes between 2019 and 2020 (at the beginning of COVID-19 pandemic), with relative differences from 0.07% to 3.17% at annual scale and 0.2% to 5.63% at quarterly scale. Our estimates differ more (in terms of relative differences) from these official estimates for quarterly and annual changes between 2020 and 2021, a period characterized by smaller changes, with relative differences from 0.81% to 3.34% at annual scale and 0.19% to 9.94% at quarterly scale.

We also compared the monthly $CO_2$ emissions of EU27 & UK and France for four / five sectors (power, industry, ground transport, domestic aviation and residential) from 2019 to 2021, with the Kayrros Carbon Watch data and the official Citepa in Figure 7.

For domestic aviation, we found that that CM-EU emissions are lower than Kayrros both in EU27 & UK and France in 2019 (Fig. 7). This is because our definition of domestic flights for France includes only metropolitan-France (mainland) trips whereas Kayrros includes France to EU, and metropolitan-France to overseas French territories emissions. This explains why Kayrros' estimates are larger. The same is true for EU27 & UK. Nevertheless, Kayrros emissions for domestic aviation are lower than CM-EU after the COVID-19 pandemic outbreak. This may be because a passenger load factor was considered for the estimates of Kayrros but not in CM-EU. The passenger load factors of flights have dropped significantly since the COVID-19 pandemic. The reason why Citepa emissions are higher than CM-EU and Kayrros for domestic aviation (Fig. 7) is because the definition of domestic flights for Citepa is all flights in and out of metropolitan-France (mainland), and half of flights between Mainland and French overseas territories (the other half is considered of the responsibility/accounting of the overseas territories). Non-commercial flights are also considered for the estimates of Citepa, which are not in CM-EU and Kayrros.

For industry, the CM-EU emissions are higher than Kayrros, but lower than Citepa. This is because the scope of Kayrros covering the regulated industrial installations is lower than the total industry sector of CM-EU, and there are the balances of attributions between power and industry for power production on industrial sites in Kayrros. While all the manufacturing industries, construction and cement production are considered in CM-EU and Citepa. Citepa also considers other industrial productions.

For power, the CM-EU emissions in EU27 & UK are almost the same as Kayrros but small differences in winter peak and summer trough (Fig. 7). This is because both datasets use the electricity production by fuel type as activity data from the same data sources. Kayrros can catch the differences from fuel burning efficiency in facility level while CM-EU can't. But the differences canceled each other when aggregating data for all countries, though they used different annual power emissions as baseline and different emission factors. In France, CM-EU only consider the emissions from power generation in the ‘power’ sector. While urban heating and other electricity self-producers industries are considered in Citepa. Thus, Citepa emissions for ‘power’ are larger than CM-EU. While Kayrros used the annual power emissions from individual French ETS plants annual reporting as the baseline, which are nearly twice larger than those of CM-EU, with largest differences during the cold season. Kayrros also considered the fuel burning efficiency in facility level. This indicates that the aggregate ‘emission factor’ based on 2019 data from EDGAR in CM-EU is lower than the emission factors deduced by Kayrros from ETS declared plant level emissions, a difference that deserves further investigation in the future. Thus, the data of Kayrros have almost the same temporal pattern but are larger with the CM-EU.

For ground transport, the CM-EU emissions are almost the same than those of the Citepa, but the Kayrros emissions are larger in the summer and lower in the winter. The mobility from trains, expected to increase in the vacation period, was not filtered from vehicles in Kayrros, so the data in the period from July to September were removed.

For residential emissions, the CM-EU emissions are higher in the summer than Citepa. This may be because we assume that the residential emissions in the summer are equal to the emissions from cooking, and cooking emissions were assumed to remain stable all the year, while the emissions from heating were assumed to depend on and vary by the heating demand.

**Technical validation for ground transportation sector and residential sector**

For technical validation, the CM-EU near-real-time daily activity data used for the ground transport sector, we compared our estimates to annual traffic counts and $CO_2$ emissions or fossil fuel use of ground transport sector for EU countries and the United Kingdom from 2010 to 2019 in Figure 8. The annual traffic counts data we used come from Eurostat[36], defined as motor vehicle movements on national territory (irrespective of registration country), covering 26 EU countries (Belgium, Bulgaria, Czech Republic, Denmark, Germany, Estonia, Ireland, Spain, France, Croatia, Italy, Cyprus, Latvia, Lithuania, Luxembourg, Hungary, Malta, Netherlands, Austria, Poland, Portugal, Romania, Slovenia, Slovakia, Finland, Sweden) and the United Kingdom. The annual fossil fuel use of ground transport sector data come from IEA[37], including five types of fossil fuel (coal, crude, oil, natural gas and peat). The $CO_2$ emissions data of ground transport are from EDGAR[8], including road transportation no resuspension, rail transportation, inland navigation and other transportation. The comparison statistics shown in Figure 8 indicate that the coefficient of determination ($R^2$) values are 0.7891 between traffic counts and CM-EU ground transport emissions and 0.7834 between traffic counts and fossil fuel use of ground transport, respectively.

For the residential sector, we compared daily emissions from CM-EU with estimates derived from natural gas use from ENSTO-G[38] in Germany, France, Italy, Poland, Netherlands, Belgium, Hungary, Romania and Luxembourg. Here we use natural gas use data from ENTSO-G as activity data instead, and annual residential emissions in 2019 from EDGAR as baseline. The average yearly emission factor of the whole residential sector is assumed to remain constant at its 2019 value, so that daily emissions are estimated as:

$$Emis_{residential,daily,2019} = Emis_{residential,yearly,2019} \times \frac{AD_{residential,daily,2019}}{AD_{residential,yearly,2019}}$$

$$Emis_{residential,daily,2020\ or\ 2021} = Emis_{residential,daily,2019} \times \frac{AD_{residential,daily,2020\ or\ 2021}}{AD_{residential,daily,2019}}$$

For comparisons, we collected daily natural gas pipeline flow data for EU27 & UK from ENTSO-G. The ENTSO-G consumption data was completed with Trading Hub Europe (THE)[39] for the German and e-control[40] for Austria. Note that the consumption sectors are not provided by e-control dataset. We further split the consumption into more detailed sectors, including household and public buildings heating, industry, and others based on energy balance datasets from Eurostat[41] following the method from Zhou et al.[42] Note that the consumption for the power sector was refined based on ENTSO-E (Carbon monitor data) as it provides higher temporal resolutions. The activity data used here are only from gas fuel and we assume that there is a linear relationship between gas fuel use and total fossil fuel use for residential sector. The results in Figure 9 show that the variation of residential emissions in our study are similar to those of ENTSO-G in all countries ($R^2$ ranges from 0.5 to 0.56), but ENTSO-G has a more prenounced winter peak and summer trough.

**Uncertainty analysis**

There are two main sources of uncertainties in the CM-EU data. 1. The uncertainty inherited from the EDGAR annual national emissions used for the reference year 2019. 2. Uncertainty from daily activity data and models used to downscale them into daily emissions. The uncertainty analysis was conducted based on the 2006 IPCC Guidelines for National Greenhouse Gas Inventories [26]. First, uncertainties were calculated for each sector based on the global Carbon Monitor methodology described in ref. [21].

For the power sector, CM-EU uses daily statistics of actual thermal production as activity data. When no uncertainty information is available, the 2-sigma uncertainty of the power activity data is assumed to be ±5% according to the IPCC recommended default uncertainty range of energy statistics[26]. In addition, for emission factors, the uncertainties mainly come from the variability of coal emission factors (as coal has a wide range of emission factors of different coal types) and changes in the mix of fuels in thermal power production. CM-EU calculates emission factors based on annual thermal production [7] and annual power emissions [43], and the uncertainty range is ±13%. We used error propagation equations to combine the aforementioned uncertainties of each part and estimated the uncertainties of annual power emissions as ±10%.

For the industry sector, a 2-sigma uncertainty (±36%) of $CO_2$ emissions from industry and cement production is estimated from monthly production data and sectoral emission factors. The uncertainty of industrial output data is assumed to be ±20% in the industry sector [44]. For the sectoral emission factor uncertainty, according to Carbon Monitor methodology, we calculate national emission factors in 2010-2012 in USA, France, Japan, Brazil, Germany, and Italy according to data availability of monthly emission data and IPI data, and their 2-sigma uncertainties vary from ±14% to ±28. Thus, we adopt a conservative uncertainty of ±30% for CM-EU emissions in this sector.

For the ground transport sector, the global Carbon Monitor methodology assessed a 2-sigma uncertainty of ±9.3% from the prediction interval of the regression model built in Paris to estimate the emissions from this sector. Note that the regression model in Paris between car counts and the TomTom congestion index was based on assuming a relative magnitude in car counts; thus, emissions follow a similar relationship with the TomTom congestion index in Paris.

For the residential sector, global Carbon Monitor compares the estimates by using our methodology with estimates from publicly available natural gas daily consumption data by residential and commercial buildings for France (https://www.smart.grtgaz.com/fr/consommation). The 2-sigma uncertainty of the daily emission estimations is further quantified as ±40%.

For the aviation sector, global Carbon Monitor compares estimates by using two different activity data, i.e., the flight route distance (what we used in this study) and the number of flights and calculate the average difference to quantify the uncertainty of ±10.2% in the aviation sector.

Overall, the uncertainty ranges of the power, ground transportation, industry, residential, and aviation are ±10.0%, ±9.3%, ±30.0%, ±40.0%, and ±10.2%, respectively and the uncertainty in the emission of EDGAR for 2019 is estimated as ±7.1% [45].

Then, we combine all the uncertainties by following the error propagation equation.

$$U_{total} = \frac{\sqrt{\sum U_s \times \mu_s}}{|\sum \mu_s|}$$

where $U_s$ and $\mu_s$ are the percentage and quantity (daily mean emissions) of the uncertainty of sector s, s respectively. The overall uncertainty is quantified as ±13.6%. We also make the technical validation for CM-EU. Therefore, the uncertainty is equal to the maximum value between the uncertainty ranges and the mean relative uncertainty from technical validation data. Finally, the overall uncertainty range of CM-EU is estimated as ±13.6%.

## Usage Notes

The generated datasets are available from https://doi.org/10.6084/m9.figshare.20219024.v1. We recommend loading the data with a script that can handle large datasets. Users should also note that the unit of emissions in this dataset is Mt $CO_2$. Latest updates and related information are available for view and download on our website https://eu.carbonmonitor.org/.

## Code availability

Python code for producing data for 27 EU countries and the United Kingdom in the dataset is provided at https://github.com/kepiyu/Carbon-Monitor-Europe/blob/main/CM_EU_v2.py.

## Acknowledgments

ZL acknowledge the National Natural Science Foundation of China (grant 71874097, 41921005, 72140002 and 72140002), Beijing Natural Science Foundation (JQ19032), and the Qiu Shi Science & Technologies Foundation.

## Acknowledgements

Authors acknowledged reviews comment to improving the manuscript.

## Author contributions

P.K., Z.L. and P.C. designed the research. P.K., Z.D. and B.Z. conducted the data processing. P.K. and P.C. wrote the manuscript. P.K., P.C. Z.D. and Z.L. designed the methods, and all authors contributed to data collection, discussion and analysis.

## Competing interests

The authors declare no competing interests.

# Figures & Tables

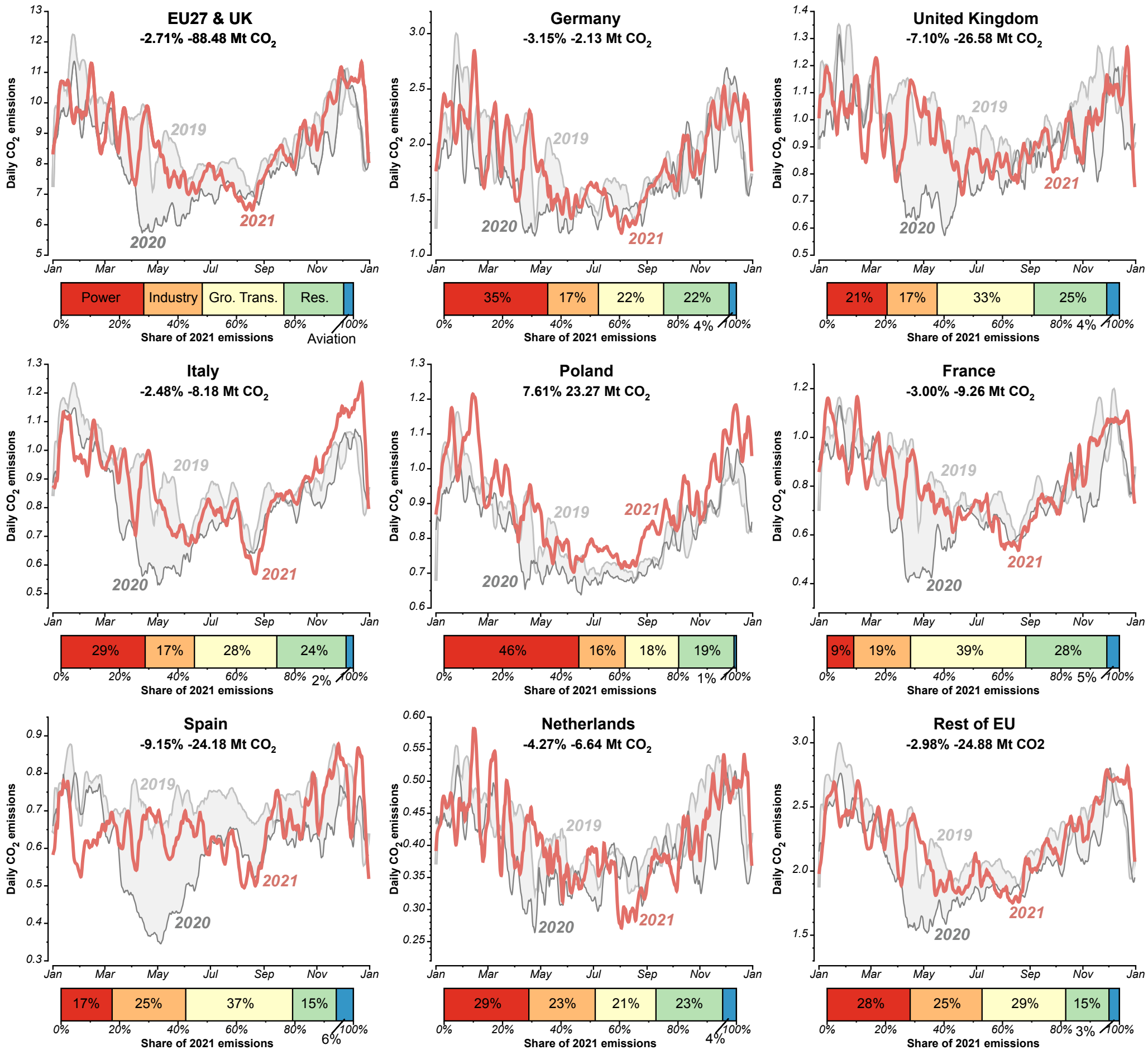

Figure 1. 7-days smoothed (running mean) daily $CO_2$ emissions for EU27+UK, Germany, the United Kingdom, Italy, Poland, France, Spain, Netherlands and the rest of European countries in 2019, 2020 and 2021. Grey shaded areas indicates the changes between 2019 and 2020. The percentage and numbers in the top of each panel reflect the relative and absolute change in 2021 compared with 2019. Bars at the bottom show the sectoral shares of annual emissions in 2021. (Aviation includes the domestic and international aviation).

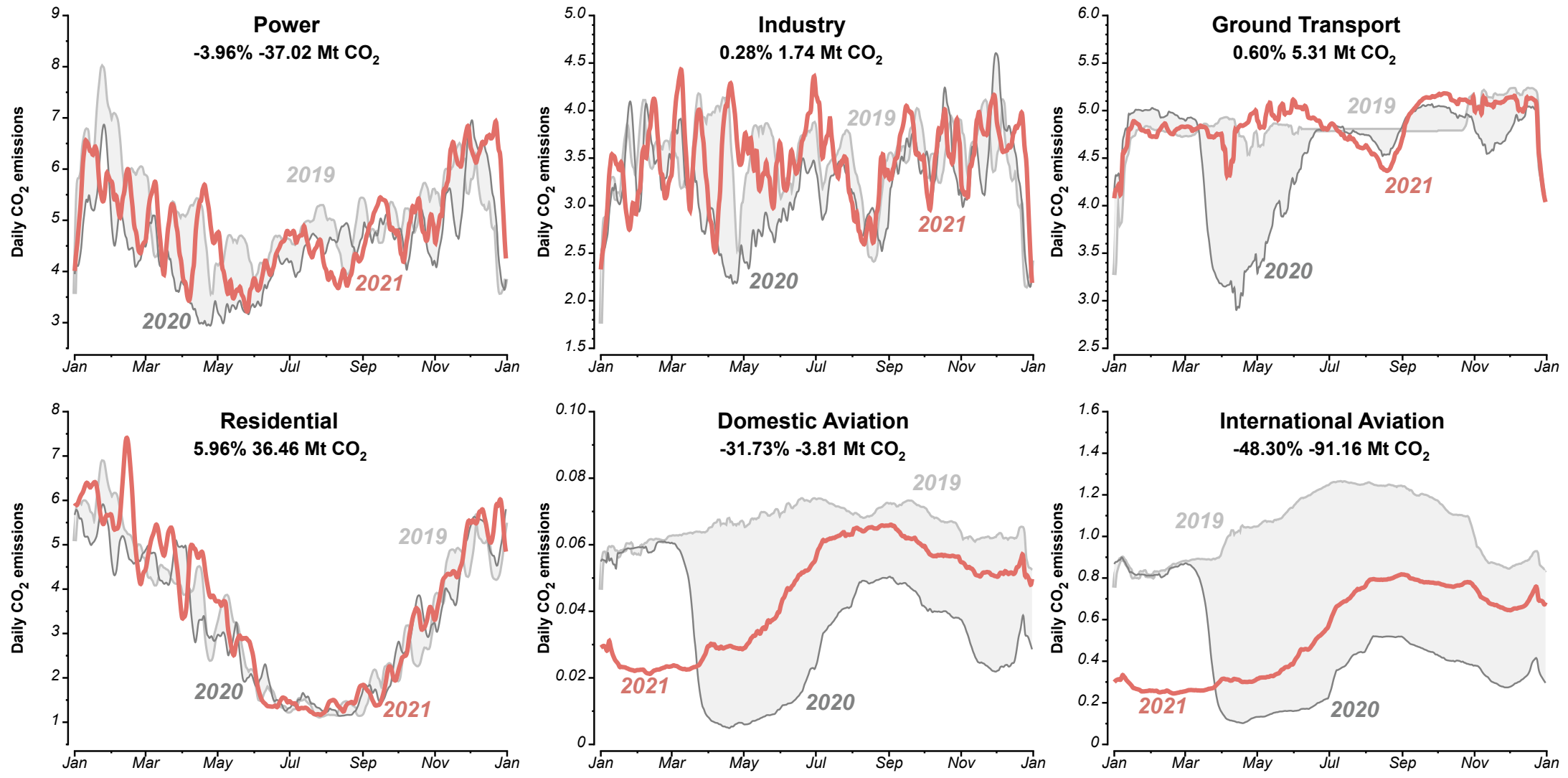

Figure 2. 7-days smoothed (running mean) daily $CO_2$ emissions for EU27 & UK for six sectors in 2019, 2020 and 2021. Grey shaded areas indicates the changes between 2019 and 2020. The percentage and number in the top of each panel reflect the relative and absolute change in 2021 compared with 2019.

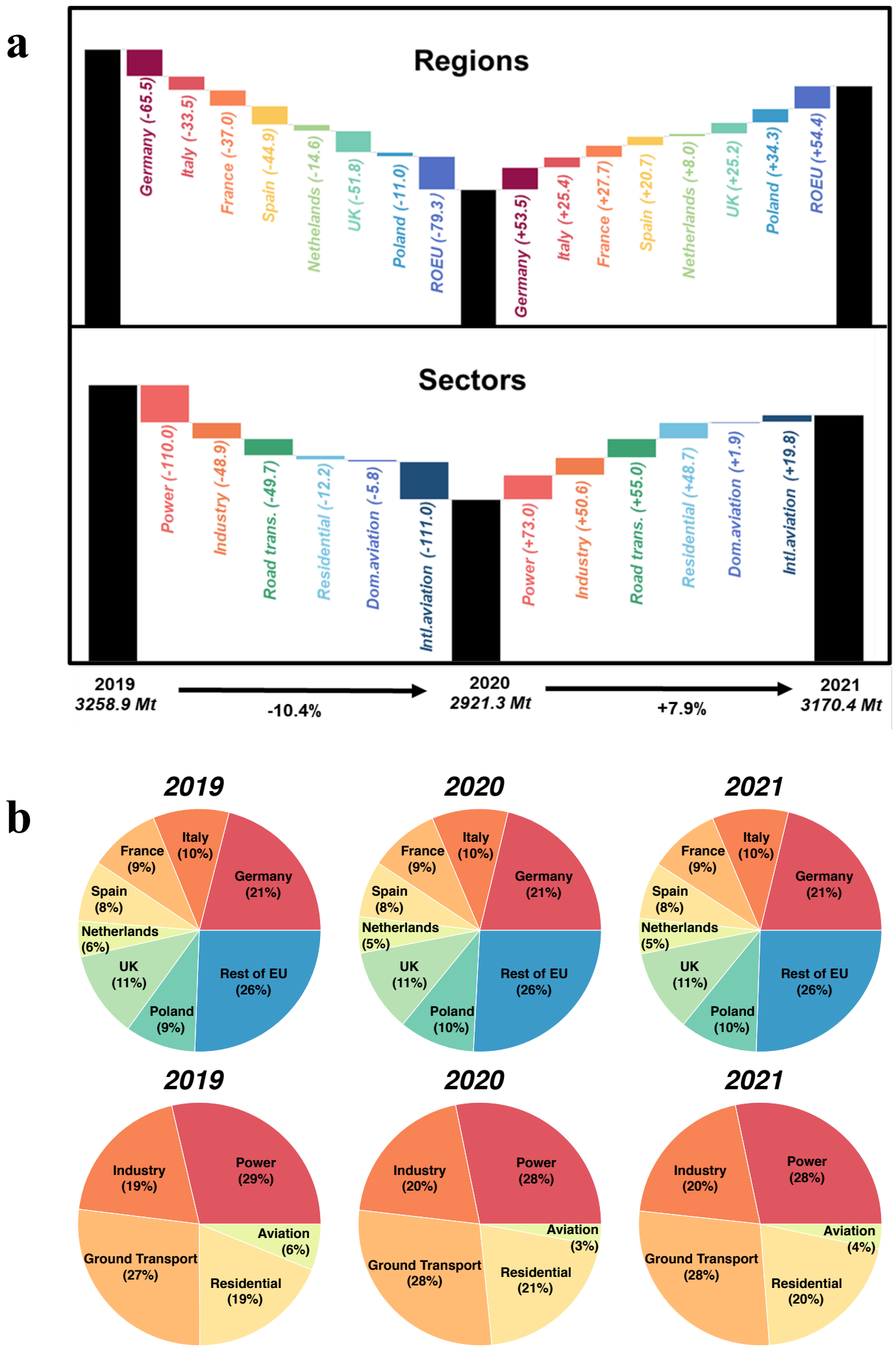

Figure 3. Changes in EU27 & UK emissions from 2019 to 2020 and 2021 across regions and sectors (a). Contribution by major countries / emitters and sectors in 2019, 2020 and 2021 (b).

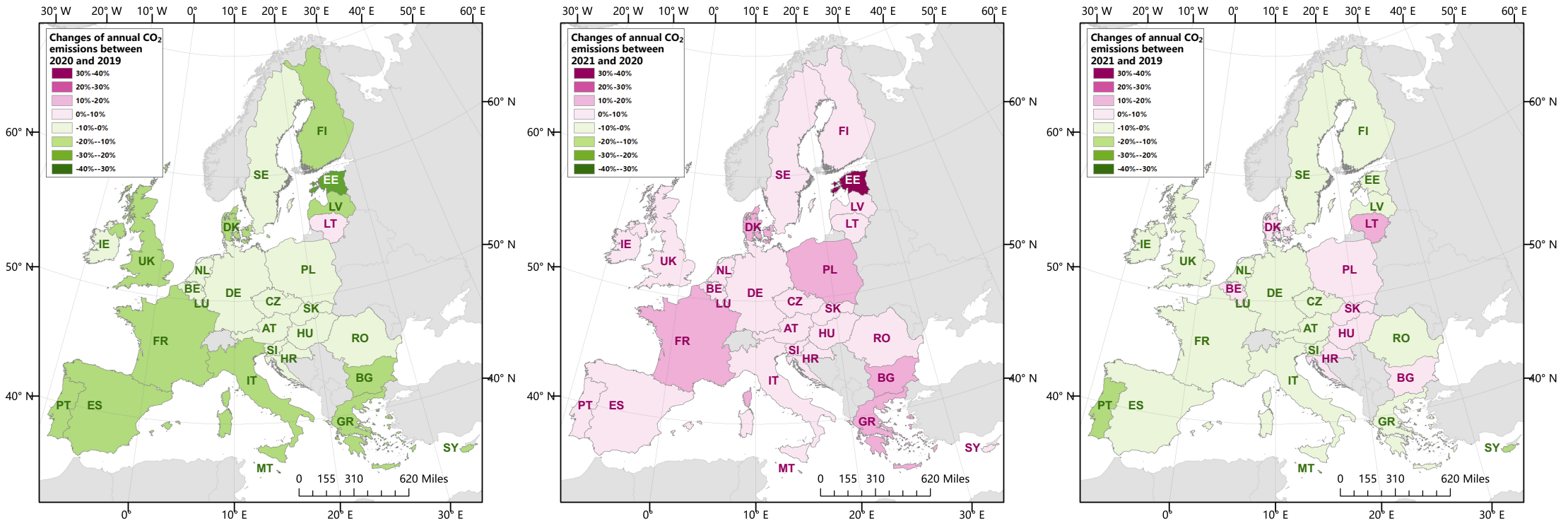

Figure 4. (left) Country-level changes in annual $CO_2$ emissions in 2020 compared with 2019, (middle) in 2021 compared with 2020, (right) in 2021 compared with 2019.

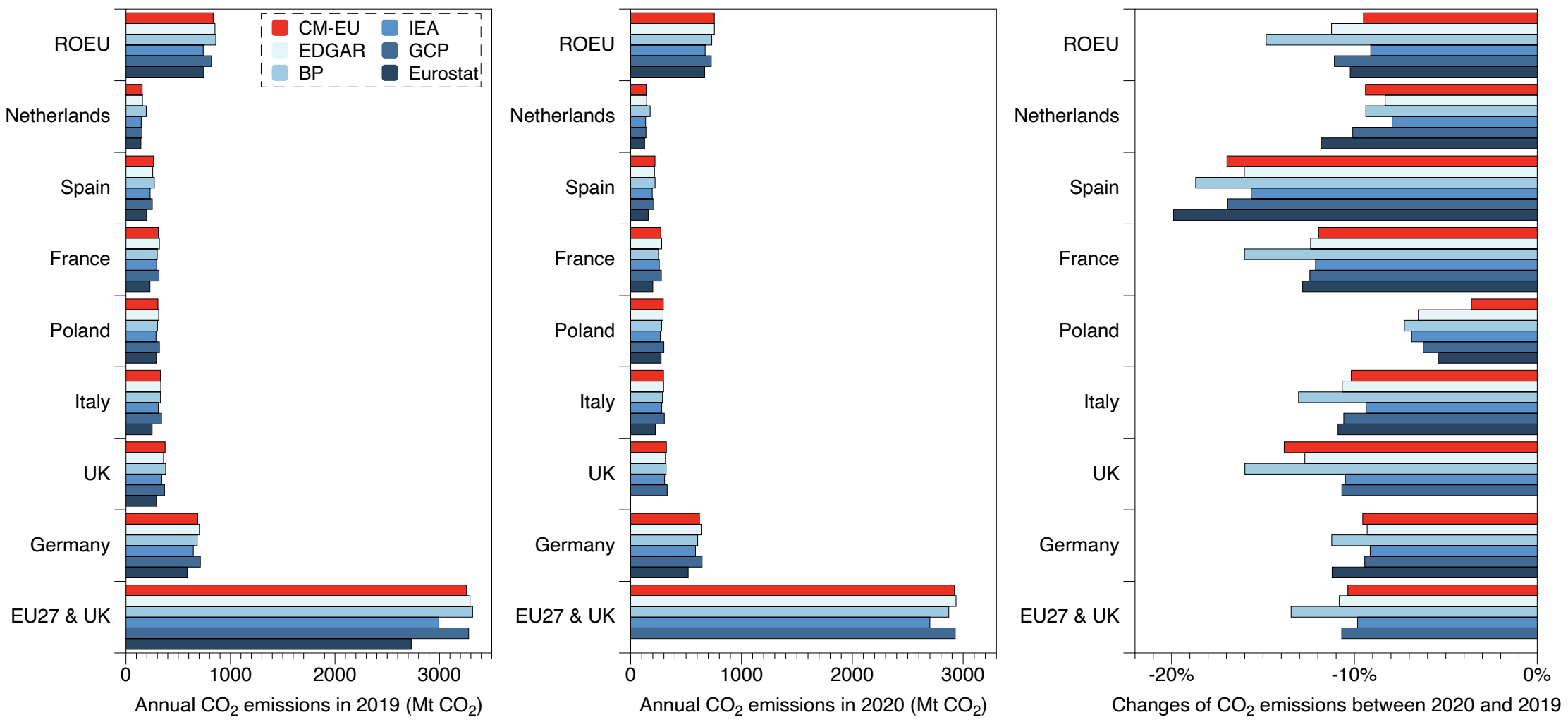

Figure 5. Comparison of annual $CO_2$ emissions in 2019 and 2020 and the relative changes between the two years among six datasets (CM-EU, EDGAR, BP, IEA, GCP and Eurostat) for EU27 & UK, Germany, UK, Italy, Poland, France, Spain, Netherlands and the rest of EU.

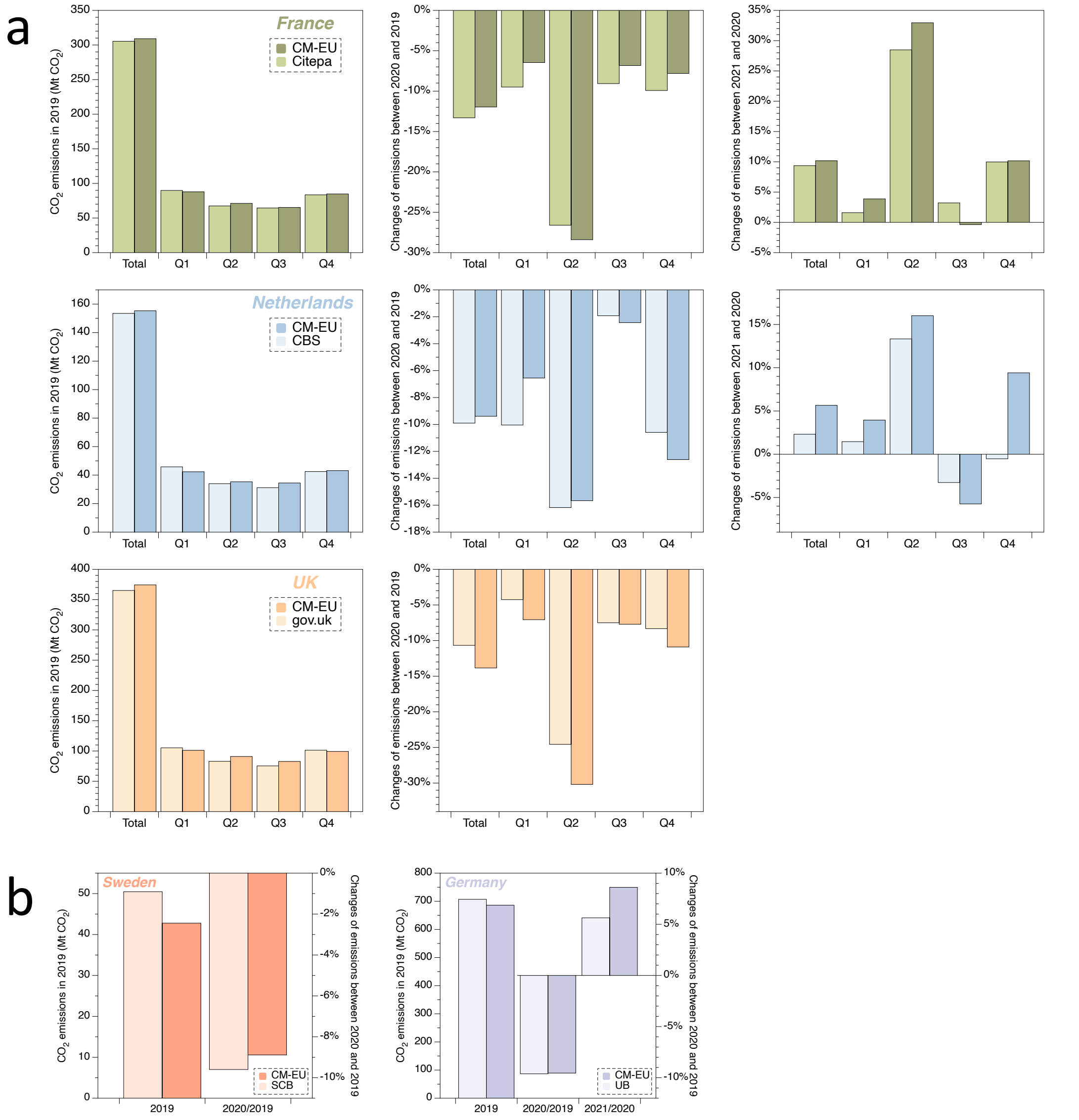

Figure 6. Comparison of quarterly $CO_2$ emissions in 2019 and relative changes between 2020 minus 2019 (and 2021 minus 2020 for France and Netherlands), UK, and annual changes for Sweden and Germany between CM-EU (deep color) and preliminary national $CO_2$ emissions estimates (light color) (quarterly in FR, NL, UK (a), annual in SW and DE (b)).

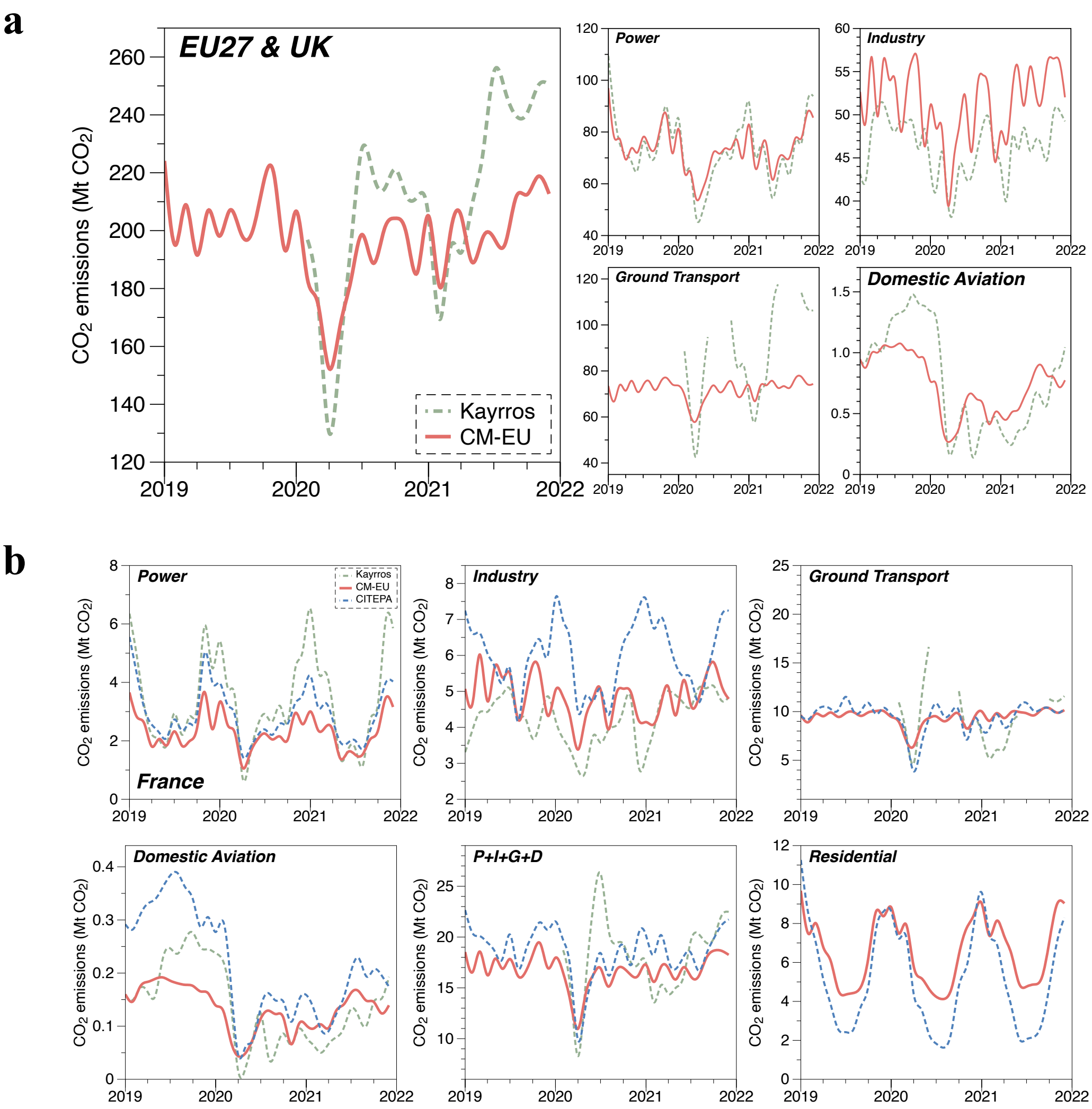

Figure 7. Technical validation of short-term emissions changes with other available 'high frequency' data. (a) Comparison of monthly $CO_2$ emissions for four sectors (power, industry, ground transport and domestic aviation) for EU27 & UK from 2019 to 2021 between CM-EU and Kayrros Carbon Watch data (see text). (b) Comparison of monthly $CO_2$ emissions in France for power, industry, ground transport, domestic aviation and residential from 2019 to 2021 between CM-EU, Kayrros Carbon Watch data[20] and Citepa monthly emission bulletin[16] (P+I+G+D means the total emissions of power, industry, ground transport and domestic aviation).

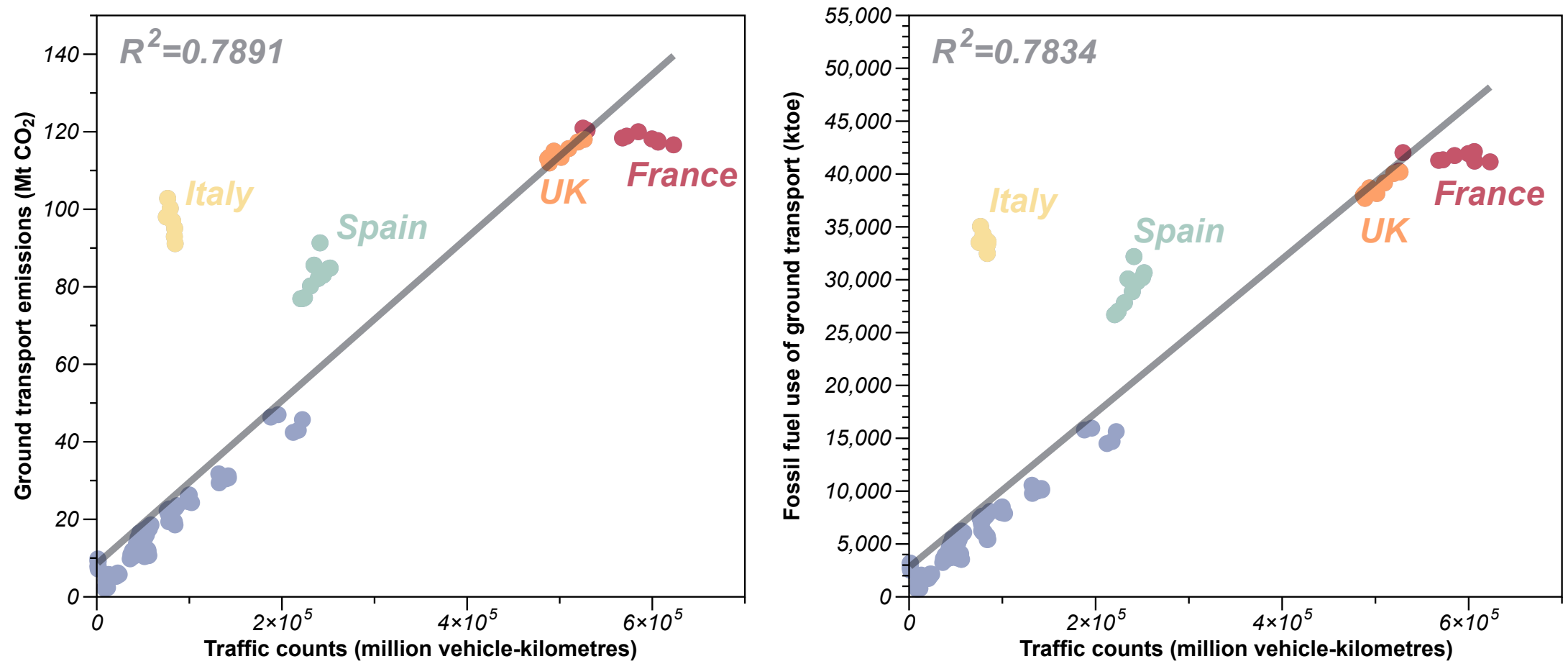

Figure 8. Scatter plots between traffic counts and $CO_2$ emissions or fossil fuel use of ground transport sector for EU countries and the United Kingdom.

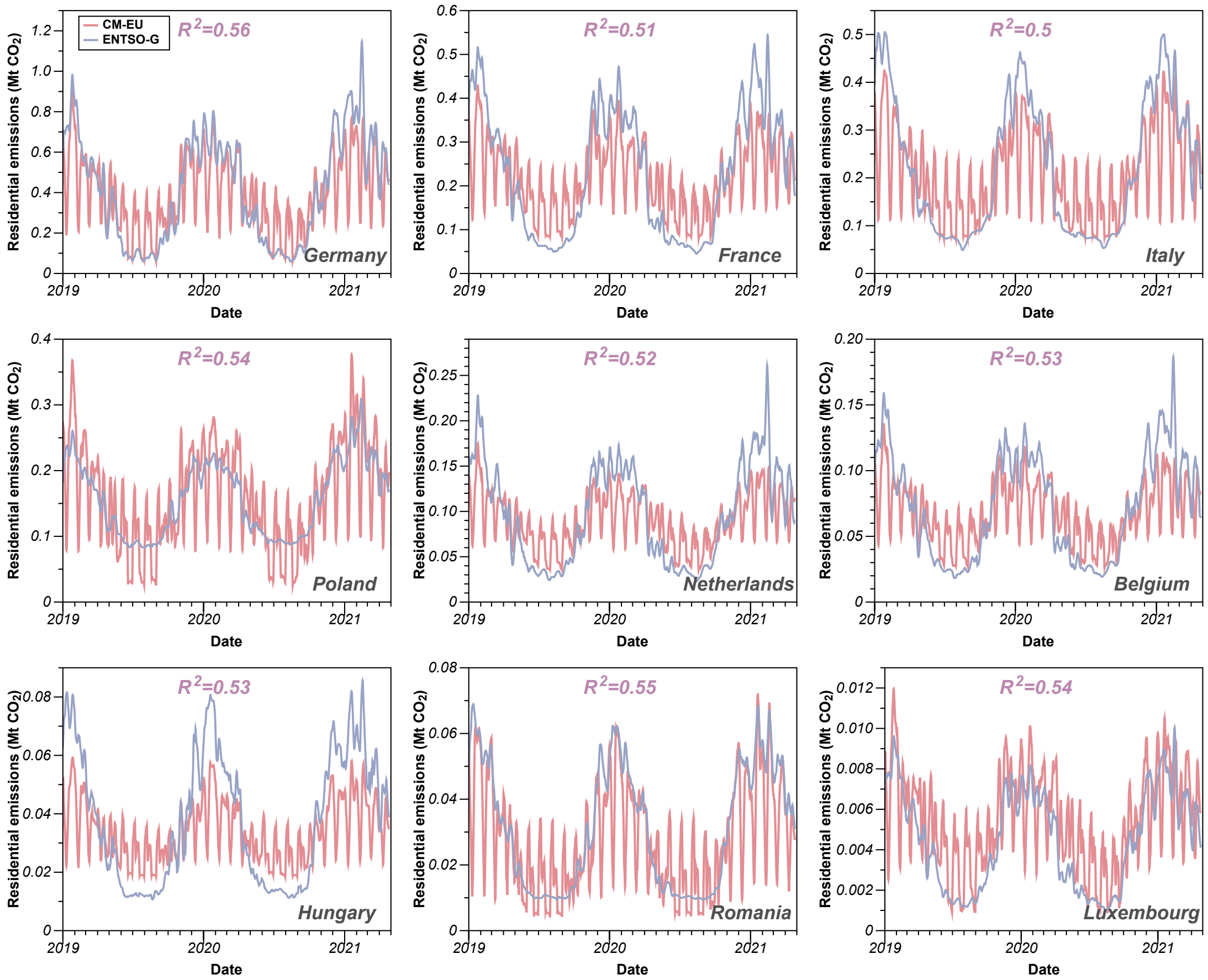

Figure 9. Comparison of daily $CO_2$ emissions of residential sector from 2019 to 2021 between CM-EU and ENTSO-G for Germany, France, Italy, Poland, Netherlands, Belgium, Hungary, Romania and Luxembourg.

| IPCC code | Sectors of EDGAR | Sectors of CM-EU |
|---|---|---|
| 1A1a | Public electricity and heat production | Power |
| 1A1bc | Other energy industries | Industry (incl. cement production) |
| 1A2 | Manufacturing industries and construction | |
| 2A1 | Cement production | |
| 1A3a | Domestic aviation | Domestic aviation |
| 1A3b | Road transportation no resuspension | Ground transportation |
| 1A3c | Rail transportation | |
| 1A3d | Inland navigation | |
| 1A3e | Other transportation | |
| 1A4 | Residential and other sectors | Residential |
| 1C1 | Memo: International aviation (bunker fuels) | International aviation |

Table 1. The mapping relationships between sectors in this study and EDGAR sectors based on IPCC categories.

| Sector | Country | Data type | Data source |
|---|---|---|---|
| Power | EU27 | Hourly thermal production | ENTSO-E Transparency platform (https://transparency.entsoe.eu/dashboard/show) |
| | UK | Hourly power generation | Balancing Mechanism Reporting Service (BMRS) (https://www.bmreports.com/) |
| Industry | EU27 (exc. Ireland) | Industrial Production Index (IPI) | Eurostat (https://ec.europa.eu/eurostat) |
| | Ireland | Industrial Production Index (IPI) | Central Statistics Office (https://data.cso.ie) |
| | UK | Industrial Production Index (IPI) | Office for National Statistics (https://www.ons.gov.uk) |
| Ground Transportation | EU27 + UK | TomTom Congestion Level | TomTom Traffic Index (https://www.tomtom.com/en_gb/traffic- index/) |
| Residential | EU27 + UK | Population-weighted heating degree days | ERA5 reanalysis of 2-meters air temperature (Copernicus Climate Change Service (C3S), 2019) |
| Aviation | EU27 + UK | Flight distance | FlightRadar24 (https://www.flightradar24.com/) |

Table 2. Sources of activity data for different sectors.

| Country/Region | Cities with TomTom data |
|---|---|
| Austria (5) | Vienna, Salzburg, Graz, Innsbruck, Linz |
| Belgium (10) | Brussels, Antwerp, Namur, Leuven, Ghent, Liege, Kortrijk, Mons, Bruges, Charleroi |
| Bulgaria (1) | Sofia |
| Czech Republic (3) | Brno, Prague, Ostrava |
| Denmark (3) | Copenhagen, Aarhus, Odense |
| Estonia (1) | Tallinn |
| Finland (3) | Helsinki, Turku, Tampere |
| France (25) | Paris, Marseille, Bordeaux, Nice, Grenoble, Lyon, Toulon, Toulouse, Montpellier, Nantes, Strasbourg, Lille, Clermont-Ferrand, Brest, Rennes, Rouen, Le Havre, Saint Etienne, Nancy, Avignon, Orleans, Le Mans, Dijon, Reims, Tours |
| Germany (26) | Hamburg, Berlin, Nuremberg, Bremen, Stuttgart, Munich, Bonn, Frankfurt am main, Dresden, Cologne, Wiesbaden, Ruhr Region West, Leipzig, Hannover, Kiel, Freiburg, Dusseldorf, Karlsruhe, Ruhr Region East, Munster, Augsburg, Monchengladbach, Mannheim, Bielefeld, Wuppertal, Kassel |
| Greece (2) | Athens, Thessaloniki |
| Hungary (1) | Budapest |
| Ireland (3) | Dublin, Cork, Limerick |
| Italy (25) | Rome, Palermo, Messina, Genoa, Naples, Milan, Catania, Bari, Reggio Calabria, Bologna, Florence, Turin, Prato, Cagliari, Pescara, Livorno, Trieste, Verona, Taranto, Reggio Emilia, Ravenna, Padua, Parma, Modena, Brescia |
| Latvia (1) | Riga |
| Lithuania (1) | Vilnius |
| Luxembourg (1) | Luxembourg |
| Netherlands (17) | The Hague, Haarlem, Leiden, Arnhem, Amsterdam, Rotterdam, Nijmegen, Groningen, Eindhoven, Utrecht, Amersfoort, Tilburg, Breda, Apeldoorn, Zwolle, Den Bosch, Almere |
| Poland (12) | Lodz, Krakow, Poznan, Warsaw, Wroclaw, Bydgoszcz, Gdansk-Gdynia-Sopot, Szczecin, Lublin, Bialystok, Bielsko-Biala, Katowice urban area |
| Portugal (5) | Lisbon, Porto, Funchal, Braga, Coimbra |
| Romania (1) | Bucharest |
| Slovakia (2) | Bratislava, Kosice |
| Slovenia (1) | Ljubljana |
| Spain (25) | Barcelona, Palma de Mallorca, Granada, Madrid, Santa Cruz de Tenerife, Seville, A Coruna, Valencia, Malaga, Murcia, Las Palmas, Alicante, Santander, Pamplona, Gijon, Cordoba, Zaragoza, Vitoria Gasteiz, Vigo, Cartagena, Valladolid, Bilbao, Oviedo, San Sebastian, Cadiz |
| Sweden (4) | Stockholm, Uppsala, Gothenburg, Malmo |
| UK (25) | Edinburgh, London, Bournemouth, Hull, Belfast, Brighton and Hove, Bristol, Manchester, Leicester, Coventry, Nottingham, Cardiff, Birmingham, Southampton, Leeds-Bradford, Liverpool, Sheffield, Swansea, Newcastle-Sunderland, Glasgow, Reading, Portsmouth, Stoke-on-Trent, Preston, Middlesbrough |

Table 3. Cities (203 across 24 EU countries and UK) where TomTom congestion level data are available.

| Dataset | Spatial coverage | Temporal coverage | Scope |
|---|---|---|---|
| EDGAR | Global | 1970-2020, annual | Fossil fuel use (combustion, flaring), industrial processes (cement, steel, chemicals and urea) and product use |
| BP | Global | 1965-2020, annual | Fossil fuel combustion |
| IEA | Global | 1960-2020, annual | Fossil fuel combustion |
| GCP | Global | 1750-2020, annual | Fossil-fuel burning, cement production, and gas flaring |
| Eurostat | European Union | 1995-2020, annual | all Nomenclature of Economic Activities (NACE activities) |
| Kayrros Carbon Watch | European Union and the UK | 2019-2021, daily | Power generation, heavy industry, ground transportation and aviation |
| Centraal Bureau van Statistiek (CBS) | Netherlands | 2019-2021, quarterly | Electricity, manufacturing, transport, agriculture, buildings and construction |
| Statistics Sweden (SCB) | Sweden | 1990-2020, annual | Territorial $CO_2$ emissions according to IPCC category |
| gov.uk | UK | 1990-2020, quarterly | Territorial $CO_2$ emissions of energy supply, business, transport, public, residential and other sectors |
| Citepa | France | 2019-2021, monthly | Territorial $CO_2$ emissions of energy, industry, agriculture and transports. |
| Umwelt Bundesamt (UB) | Germany | 1990-2021, annual | Territorial $CO_2$ emissions |

Table 4. List of $CO_2$ emissions datasets used for technical validation of CM-EU products